\documentclass{PoS}

\usepackage{amsmath}
\usepackage{caption}
\usepackage{multirow}

\newcommand{\castorc}{Computation-based Science and Technology
  Research Center, The Cyprus Institute, 20 Konstantinou Kavafi Str.,
  2121 Nicosia, Cyprus}

\newcommand{\ucy}{Department of Physics, University of Cyprus,
  P.O. Box 20537, 1678 Nicosia, Cyprus}

\newcommand{\desy}{NIC, DESY, Platanenallee 6, D-15738 Zeuthen,
  Germany}

\newcommand{\temple}{Department of Physics, Temple University,
  Philadelphia, PA}

\newcommand{\milan}{INFN Sezione di Milano-Bicocca, Milano, Italy}

\newcommand{\mainz}{Institut f\"ur Kernphysik, Johannes
  Gutenberg-Universit\"at Mainz, Mainz, Germany}

\title{Nucleon electromagnetic and axial form factors with
  N$_\textrm{f}$=2 twisted mass fermions at the physical point}

\ShortTitle{Nucleon electromagnetic and axial form factors}

\author{Constantia Alexandrou \\\ucy\, and \castorc\\  E-mail:
  \email{alexand@ucy.ac.cy}}

\author{Martha Constantinou \\\temple\\  E-mail: \email{marthac@temple.edu}}

\author{Kyriakos Hadjiyiannakou,
  Christos Kallidonis,
  \speaker{Giannis Koutsou} \\\castorc\\  E-mail:
  \email{k.hadjiyiannakou@cyi.ac.cy},
  \email{c.kallidonis@cyi.ac.cy},
  \email{g.koutsou@cyi.ac.cy}}

\author{Karl Jansen \\\desy\\  E-mail:
  \email{karl.jansen@desy.de}}

\author{Konstantin Ottnad \\\mainz\\
  E-mail:\email{kottnad@uni-mainz.de}}

\author{Alejandro Vaquero \\\milan\\
   E-mail:\email{alexvaq@physics.utah.edu}}

\abstract{We present results for the nucleon electromagnetic and axial
  form factors using an N$_\textrm{f}$=2 twisted mass fermion ensemble
  with pion mass of about 131 MeV. We use multiple sink-source
  separations to identify excited state contamination. Dipole masses
  for the momentum dependence of the form factors are extracted and
  compared to experiment, as is the nucleon magnetic moment and charge
  and magnetic radii.}

\FullConference{34th annual International Symposium on Lattice Field Theory\\  
  24-30 July 2016\\
  University of Southampton, UK}

\begin{document}

\section{Introduction}
Form factors of the nucleon are fundamental probes of its
structure. The electromagnetic form factors are related to the nucleon
magnetic moment, its electric and magnetic radii. Axial form factors
probe chiral symmetry and test partial conservation of the axial
current (PCAC), having been studied in chiral effective theories.

Both electromagnetic and axial form factors have been extensively
studied in lattice QCD. Recent experimental results, in combination
with availability of simulations with physical quark masses on the
lattice, have increased interest in an \textit{ab initio} calculation
of these form factors. These include tension between the value
obtained for the proton radius between electron
scattering~\cite{Mohr:2015ccw} and hydrogen
spectroscopy~\cite{Pohl:2010zza} as well as with recent measurements
of muonic deuterium spectroscopy~\cite{Pohl669}. Furthermore recent
re-analyses of neutrino scattering
data~\cite{AguilarArevalo:2010zc,Meyer:2016oeg} report large
systematics in the determination of the axial dipole mass $M_A$. In
this contribution, we calculate the axial and electromagnetic form
factors of the nucleon on an ensemble of twisted mass fermion
configurations with clover improvement and two degenerate light quarks
(N$_\textrm{f}=2$) tuned to reproduce a pion mass of about
131~MeV~\cite{Abdel-Rehim:2015pwa}. We use multiple sink-source
separations and $\mathcal{O}(10^5)$ statistics to evaluate excited
state effects in these quantities.

\section{Setup and lattice parameters}
\subsection{Axial and Electromagnetic form factors}
Form factors are extracted from nucleon matrix elements:
\[
\langle N(p', s')|\mathcal{O}^{X}_\mu|N(p,s)\rangle =
\sqrt{\frac{m^2_N}{E_N(\vec{p}')E_N(\vec{p})}}\bar{u}_N(p',s')\Lambda^{X}_\mu(q^2)u_N(p,s)
\]
with $N(p,s)$ a nucleon state of momentum $p$ and spin $s$,
$E_N(\vec{p}) = p_0$ its energy and $m_N$ its mass, $q=p'-p$, the
momentum transfer from initial ($p$) to final ($p'$) momentum, $u_N$ a
nucleon spinor and $\mathcal{O}^{X}$ either the axial ($X=A$) or
vector $(X=V)$ current. For the case of axial form factors, we use the
axial current: $\mathcal{O}^{A}_\mu$ = $A^3_\mu$ =
$\bar{\psi}\frac{\tau_3}{2}\gamma_5\gamma_\mu\psi$, with
$\bar{\psi}=(\bar{u}, \bar{d})$, $u$ and $d$ up- and down-quark fields
and $\tau_3$ the third Pauli matrix acting on flavor space. For the
electromagnetic form factors we use the isovector, symmetrized lattice
conserved vector current $\mathcal{O}^{V}_\mu$ =
$\frac{1}{2}[j_\mu(x)+j_\mu(x-\hat{\mu})]$, with $j_\mu(x)$ the Wilson
conserved current. Use of the isovector current means that
disconnected contributions cancel. Furthermore, use of the conserved
electromagnetic current means no renormalization of the vector
operator is required. For the axial form factors we use
$Z_A=$0.7910(4)(5)~\cite{Abdel-Rehim:2015owa}. The matrix element of
the axial current yields the axial $G_A$ and induced pseudo-scalar
$G_p$ form factors, while the vector current yields the Dirac $F_1$
and Pauli $F_2$ form factors:
\begin{align}
  \Lambda^A_\mu(q^2) = \frac{i}{2}\gamma_5\gamma_\mu G_A(q^2)+\frac{q_\mu\gamma_5}{2m_N}G_p(q^2),\quad\Lambda_\mu^V(q^2) = \gamma_\mu F_1(q^2)+\frac{i\sigma_{\mu\nu}q^\nu}{2m_N}F_2(q^2).
\end{align}
The Dirac and Pauli form factors can also be expressed in terms of the
nucleon electric $G_E$ and magnetic $G_M$ Sachs form factors via
$G_E(q^2) = F_1(q^2)+\frac{q^2}{(2m_N)^2}F_2(q^2)$ and $G_M(q^2) =
F_1(q^2)+F_2(q^2)$.

\subsection{Lattice extraction of form factors}
On the lattice, extraction of matrix elements requires calculating a
three-point correlation function. We use sequential inversions through
the sink fixing the sink momentum $\vec{p}'$ to zero, which constrains
$\vec{p}=-\vec{q}$. We form a ratio of three- to two-point functions
which, after taking the large time limit, cancel unknown overlaps and
energy exponentials: $R_\mu(\Gamma;\vec{q};t_s;t_{\rm
  ins})\xrightarrow[t_{\rm ins}\gg]{t_s-t_{\rm
    ins}\gg}\Pi_\mu(\Gamma;\vec{q})$, where $R_\mu$ is the ratio of
three- to two-point functions as defined in
Ref.~\cite{Alexandrou:2010hf}, $t_s$ ($t_\textrm{ins}$) the sink
(insertion) time assuming the source is at the origin, and $\Gamma$
the sink polarization.

In what follows we will use two methods to extract $\Pi_\mu$ from
lattice data: i) in the standard \textit{plateau} method, we fit the
$t_{\rm ins}$ dependence of $\Pi_\mu$ to a constant for multiple $t_s$
values observing the dependence with $t_s$, as shown for $G_E$ in the
left panel of Fig.~\ref{fig:plateau, summation}. Excited states are
suppressed when our result does not change with $t_s$. ii) in the
\textit{summation} method, we calculate: $\sum_{t_{\rm
    ins}}R_\mu(\Gamma;\vec{q};t_s;t_{\rm
  ins})\xrightarrow{t_s\gg}\Pi_{\mu}(\Gamma;\vec{q})t_s+C$ and carry
out a two-parameter fit for obtaining the slope, as in the right panel
of Fig.~\ref{fig:plateau, summation}.

\begin{figure}
  \begin{minipage}{0.367\linewidth}
    \includegraphics[width=\linewidth]{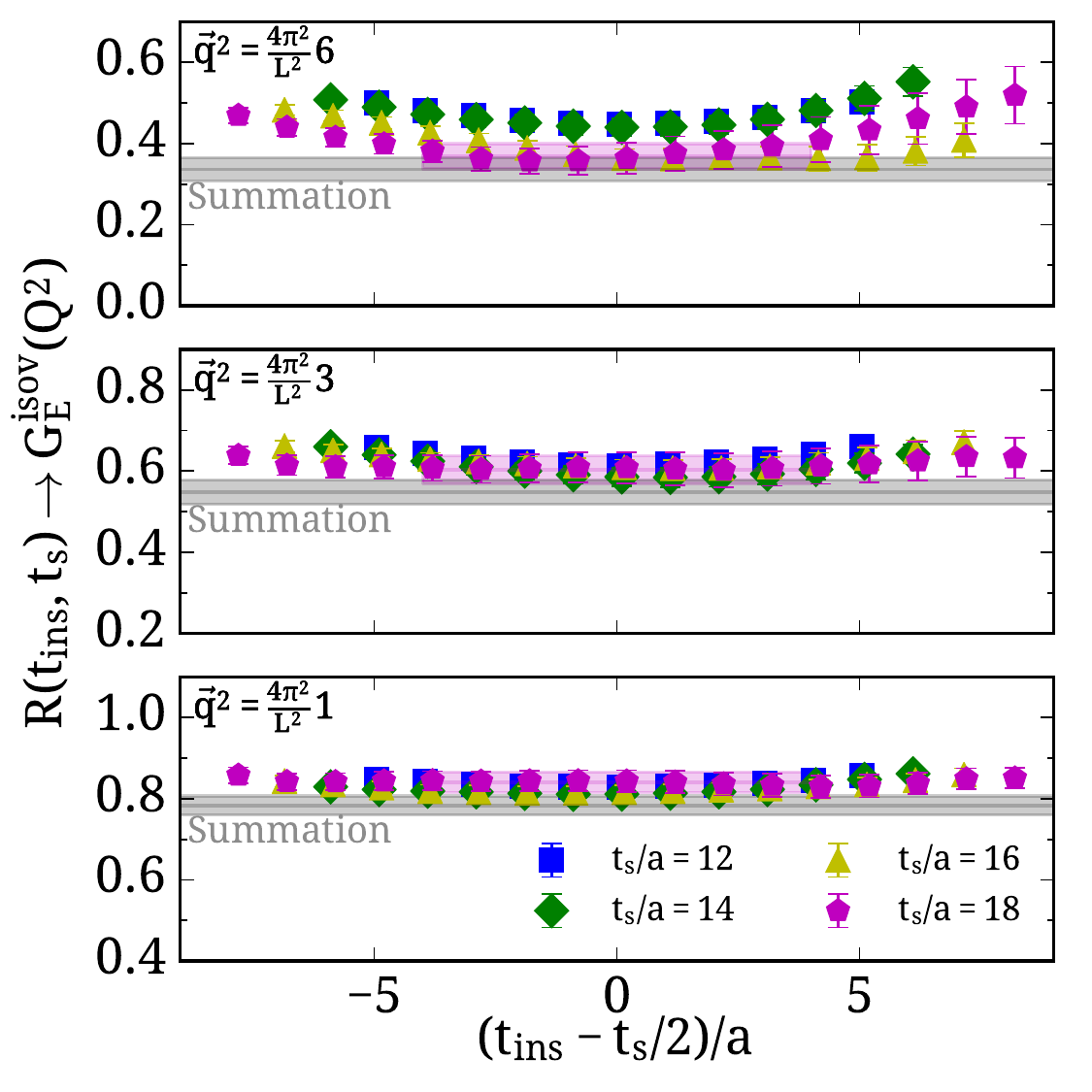}
  \end{minipage}
  \begin{minipage}{0.367\linewidth}
    \includegraphics[width=\linewidth]{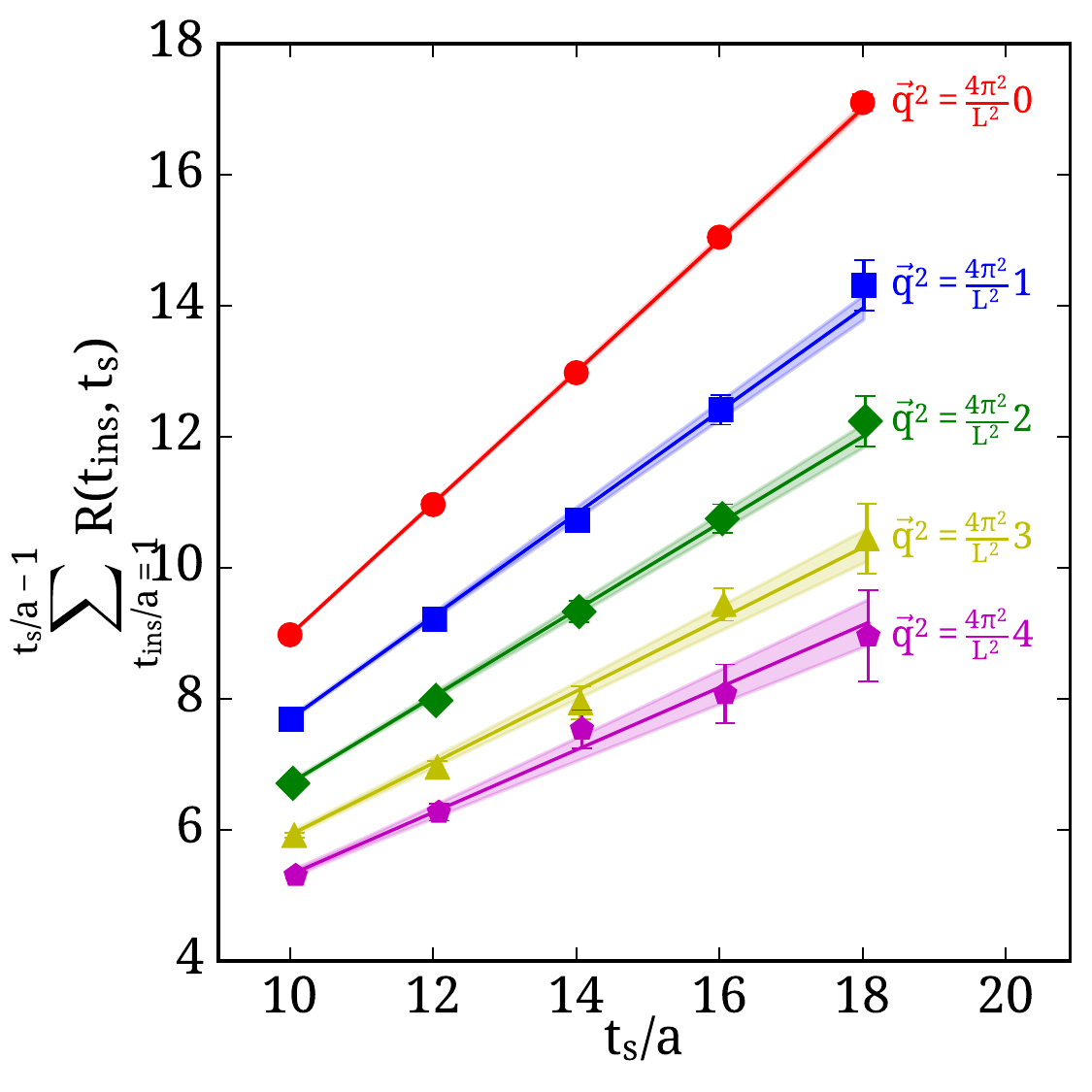}
  \end{minipage}
  \hfill  
  \begin{minipage}{0.244\linewidth}
    \caption{Example fits for $G_E$, left for the plateau method for
      three lattice momenta, and right for the summation method for
      the first five lattice momenta. On the left, with the grey
      bands, we also show the result of the summation method.}
    \label{fig:plateau, summation}    
  \end{minipage}
\end{figure}

Having $\Pi_\mu(\Gamma;\vec{q})$, different combinations of current
insertion directions ($\mu$) and nucleon polarizations determined by
$\Gamma$ yield different form factors. Using $\Pi^V$ to denote
electromagnetic and $\Pi^A$ for axial matrix elements, we have:
\begin{align}
  \Pi^V_0(\Gamma_0;\vec{q}) =&
  \mathcal{C}\frac{E_N+m_N}{2m_N}G_E(Q^2),\quad&
  \Pi^V_i(\Gamma_0;\vec{q}) =&
  \mathcal{C}\frac{q_i}{2m_N}G_E(Q^2)\label{eq:ffs}\\ \Pi^V_i(\Gamma_k;\vec{q})
  =& \mathcal{C}\frac{\epsilon_{ijk}q_j}{2m_N}G_M(Q^2),\quad&
  \Pi^A_i(\Gamma_k;\vec{q}) =&
  \frac{i\mathcal{C}}{4m_N}[\frac{q_kq_i}{2m_N}G_p(Q^2) - (E_N+m_N)\delta_{ik}G_A(Q^2)]\nonumber
\end{align}
where $Q^2=-q^2$, $\mathcal{C}=\sqrt{\frac{2m_N^2}{E_N(E_N+m_N)}}$,
the unpolarized projector: $\Gamma_0=\frac{1+\gamma_0}{4}$, the
polarized projector: $\Gamma_k=i\gamma_5\gamma_k\Gamma_0$, and
$i,k=1,2,3$.

\subsection{Lattice setup}
We use a lattice with volume 48$^3\times$96 and lattice spacing
determined at a~$\simeq$~0.093~fm~\cite{Abdel-Rehim:2016won}. The
parameters of the calculation are summarized in
Table~\ref{table:statistics}. This setup allows calculation of $G_E$
on all five sink-source separations and of $G_M$, $G_A$ and $G_p$ on
the three smallest. $G_E$ and $G_M$ can be extracted directly via
Eq.~(\ref{eq:ffs}) since they depend on different sink
projectors. $G_A$ and $G_p$ are both extracted from the last
expression of Eq.~(\ref{eq:ffs}). We separate the two form factors via
an over-constrained fit, solving the resulting eigenvalue problem via
singular value decomposition, as explained in
Ref.~\cite{Alexandrou:2007xj}.
\begin{table}
  \begin{minipage}{0.45\linewidth}
    \begin{tabular}{ccr@{ = }r}
      \hline\hline
      $t_s$ [a] & Proj. & $N_{\rm cnf}\cdot N_{\rm src}$&$N_{\rm st}$ \\
      \hline
      10,12,14 & $\Gamma_0$, $\Gamma_k$ & 578$\times$16& 9248 \\
      16 & $\Gamma_0$ & 530$\times$88& 46640 \\
      18 & $\Gamma_0$ & 725$\times$88& 63800 \\
      \hline\hline
    \end{tabular}
  \end{minipage}
  \hfill
  \begin{minipage}{0.55\linewidth}
    \caption{Form factor calculation setup. The first column shows the
      sink-source separations used, the second column the sink
      projectors and the last column the total statistics ($N_{\rm
        st}$) obtained using $N_{\rm cnf}$ configurations times
      $N_{\rm src}$ source-positions per configuration.}
    \label{table:statistics}
  \end{minipage}
\end{table}

\section{Results}
\subsection{Axial form factors}

\begin{figure}
  \begin{minipage}[t]{0.49\linewidth}
    \includegraphics[width=\linewidth]{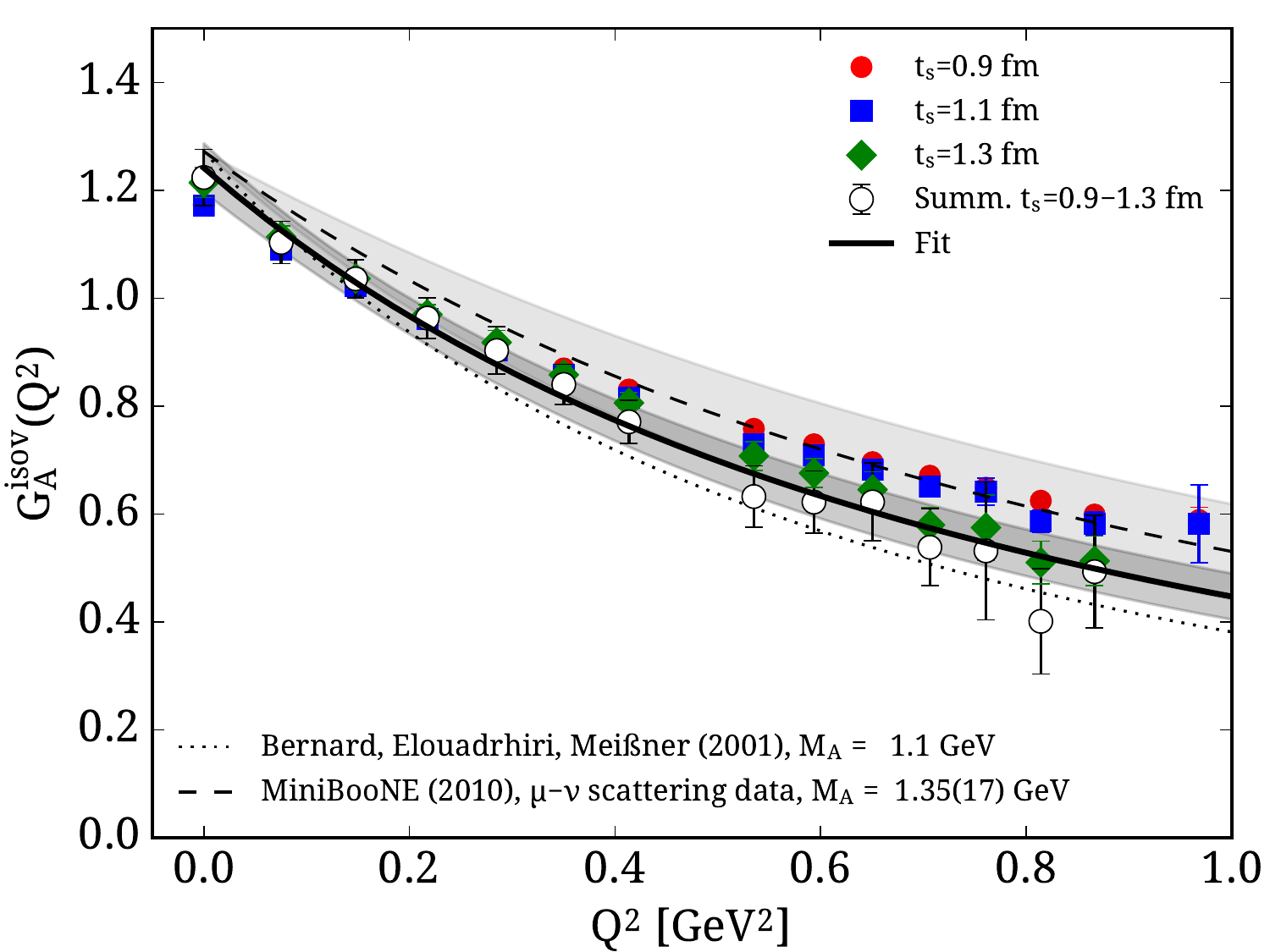}\\[3ex]
    \includegraphics[width=\linewidth]{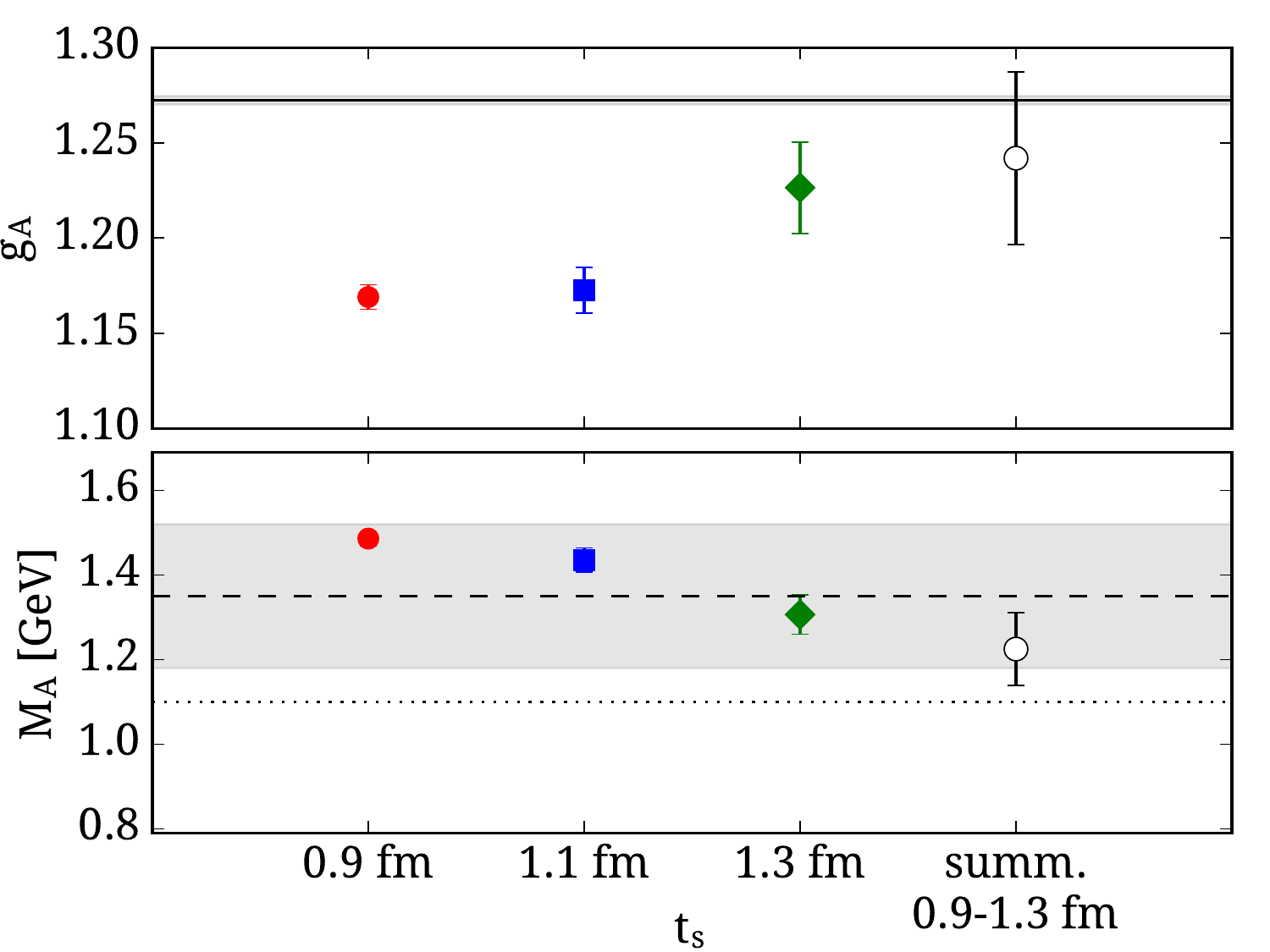}
  \end{minipage}
  \hfill
  \begin{minipage}[t]{0.49\linewidth}
    \includegraphics[width=\linewidth]{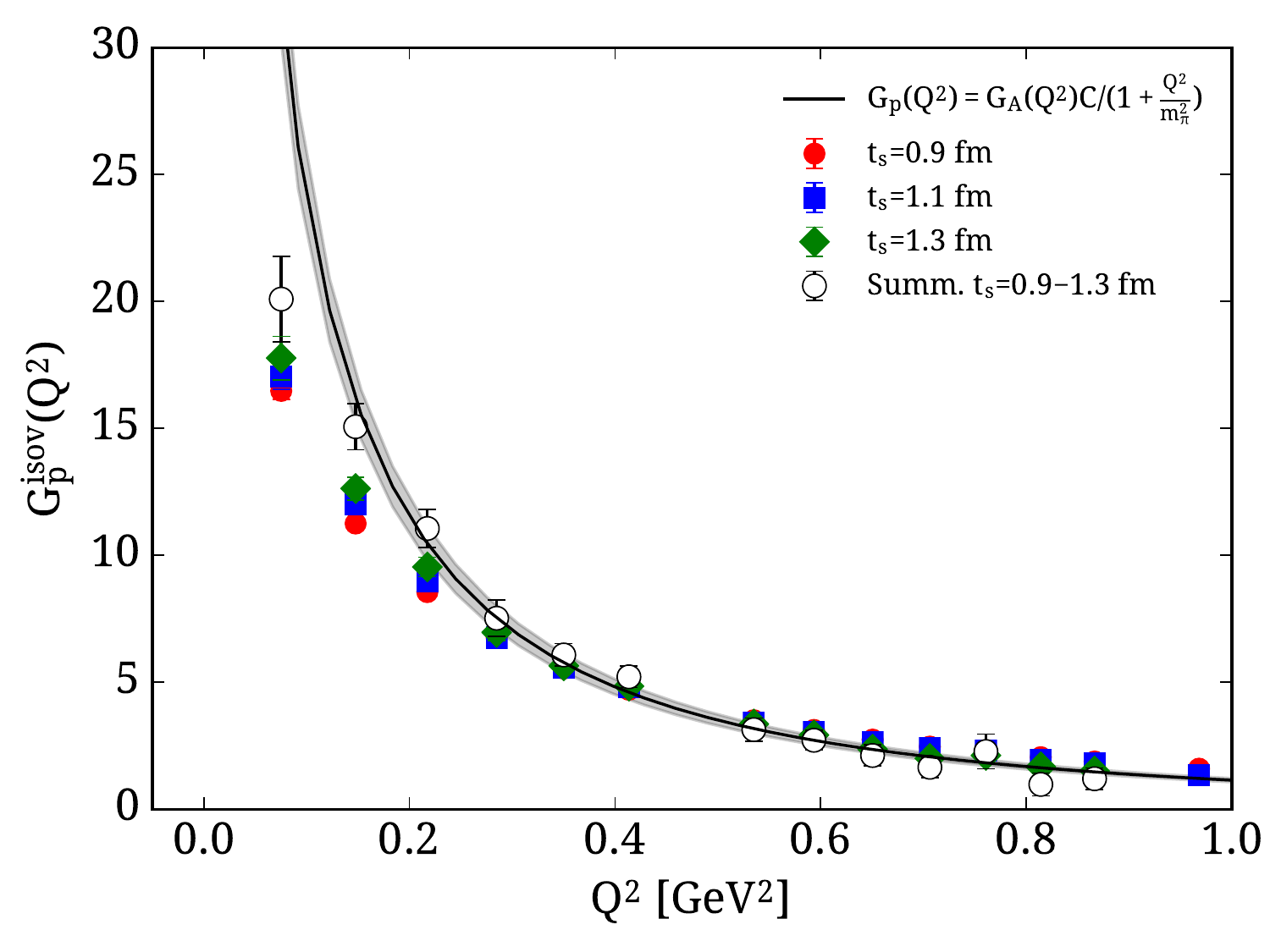}
    \caption{Left: Axial nucleon form factor using $t_s\simeq$ 0.9 fm
      (red circles), 1.1 fm (blue squares), and 1.3 fm (green
      diamonds) and using the summation method (open circles). The
      solid line (upper panel) is a fit of the latter to a dipole
      form. The dashed and dotted lines are from
      Refs.~\protect\cite{AguilarArevalo:2010zc} and~\protect\cite{Bernard:2001rs}
      respectively. Dipole fit results (lower panel) are compared with
      $g_A$ from Ref.~\protect\cite{Olive:2016xmw} shown with the solid
      line. Top: Results for the induced pseudo-scalar form factor
      $G_p$ with the notation of the left panel.}
    \label{fig:axial ffs}
  \end{minipage}
\end{figure}

The axial form factors are shown in Fig.~\ref{fig:axial ffs}. For
$G_A$ we see values increasing at low $Q^2$ as the sink-source
separation increases, while at larger momenta we see a decreasing
trend. We fit all three sink-source separations, and the summation
method, to a dipole form: $G_A(Q^2)=\frac{g_A}{(1+Q^2/M_A^2)^2}$,
allowing both the axial charge $g_A$ and the axial mass $M_A$ to vary.
We observe $g_A$ approaching its experimental value with increasing
sink-source separation. More details on this calculation of $g_A$ can
be found in Ref.~\cite{Alexandrou:2016pos} in these proceedings. $M_A$
is found consistent within errors of a recent experimental
determination~\cite{AguilarArevalo:2010zc} shown with the dashed line
in the central panel of Fig.~\ref{fig:axial ffs}. We note that our
values of $M_A$ are also consistent within the wide error of a recent
reanalysis of experimental data, not shown in Fig.~\ref{fig:axial
  ffs}, which yields $M_A$=1.01(24)~GeV~\cite{Meyer:2016oeg}.

The induced pseudo-scalar form factor $G_p$ exhibits similar
excited-state dependence at low $Q^2$. Assuming a pion-pole motivates
the form: $G_p(Q^2) = G_A(Q^2)C/(1+\frac{Q^2}{m_\pi^2})$ to which we
fit to using a dipole form for $G_A$ thus requiring only $C$ to
vary. We obtain $\sqrt{C}/2=5.9(2)$ to be compared to the
phenomenological expectation
$\sqrt{C}/2=m_N/m_\pi=7.16(4)$~\cite{Alexandrou:2007xj}.

\subsection{Electromagnetic form factors}
\begin{figure}
  \begin{minipage}{0.49\linewidth}
    \includegraphics[width=\linewidth]{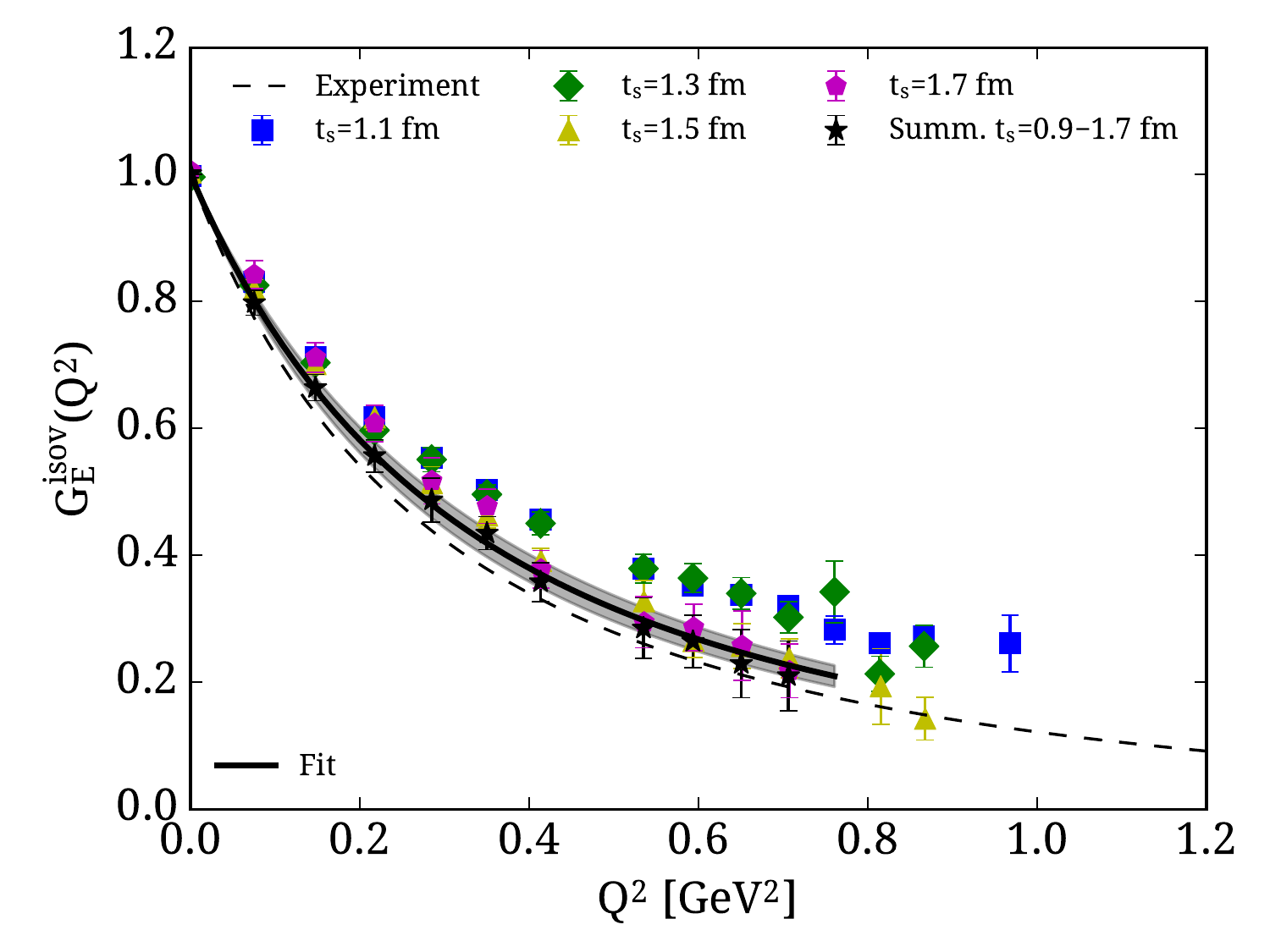}
  \end{minipage}
  \hfill
  \begin{minipage}{0.49\linewidth}
    \includegraphics[width=\linewidth]{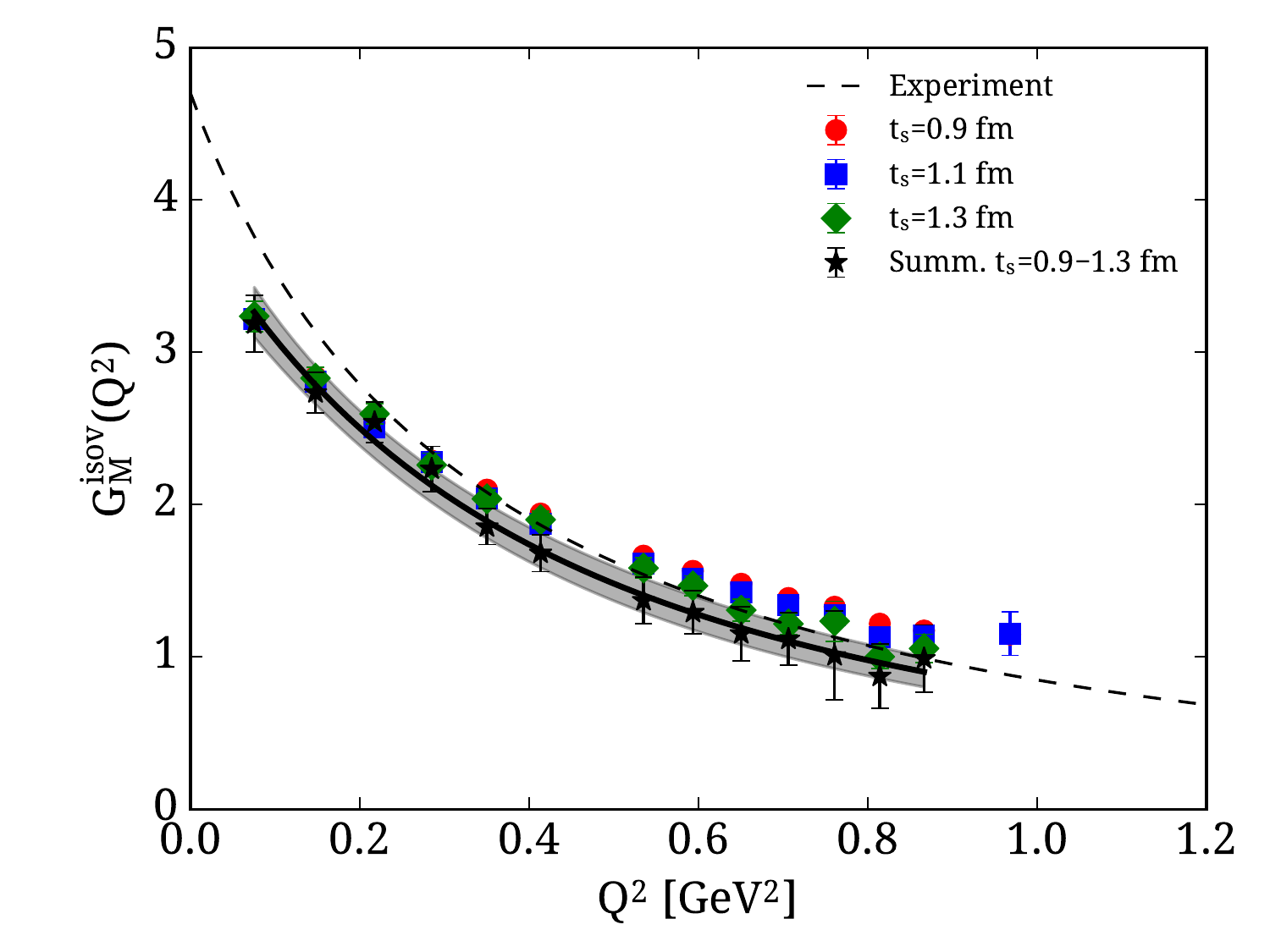}
  \end{minipage}
  \caption{Isovector electric (left) and magnetic (right) Sachs
    form factors. For $G_E$ we show with yellow triangles and magenta
    pentagons $t_s\simeq$ 1.5 and 1.7~fm respectively. Asterisks
    denote the summation method. The solid line and band denotes fits
    to a dipole form as explained in the text. The dashed line is the
    experimental parameterization.}
  \label{fig:em ffs}
\end{figure}

The isovector electromagnetic Sachs form factors are shown in
Fig.~\ref{fig:em ffs}, where for $G_E$ two additional $t_s$ values are
available. For $G_E$ we see a tendency towards the experimental
results as $t_s$ increases. The same is not observed for $G_M$ which
underestimates the low-$Q^2$ experimental values and which decreases
with increasing $t_s$.

\begin{figure}
  \begin{minipage}{0.49\linewidth}
    \includegraphics[width=\linewidth]{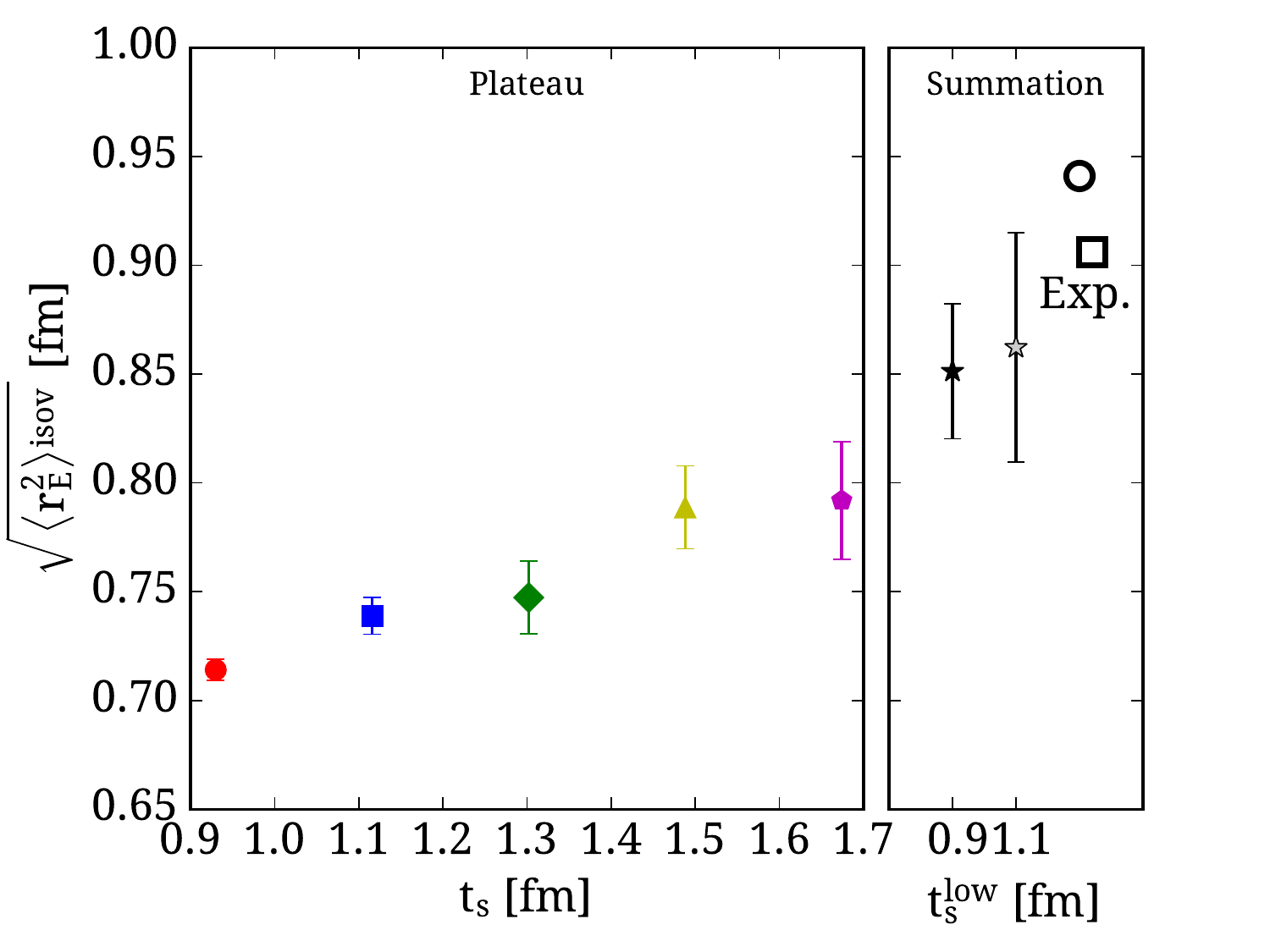}
  \end{minipage}
  \hfill
  \begin{minipage}{0.49\linewidth}
    \includegraphics[width=\linewidth]{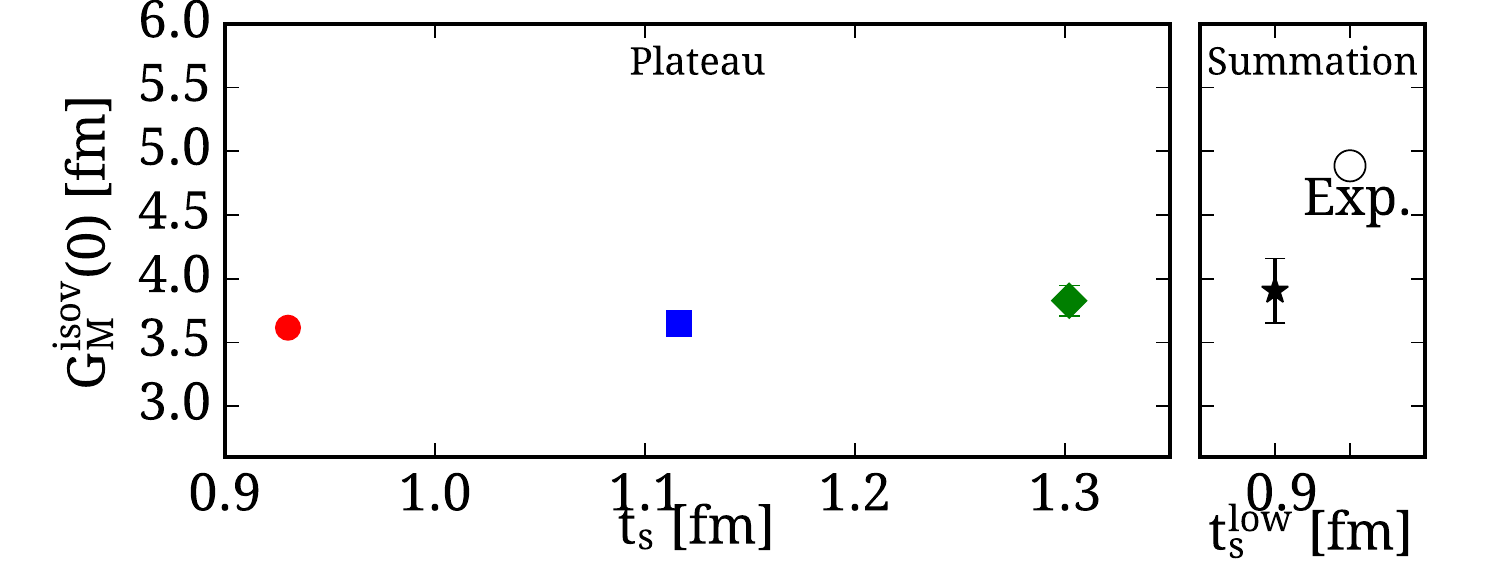}\\[-3.2ex]
    \includegraphics[width=\linewidth]{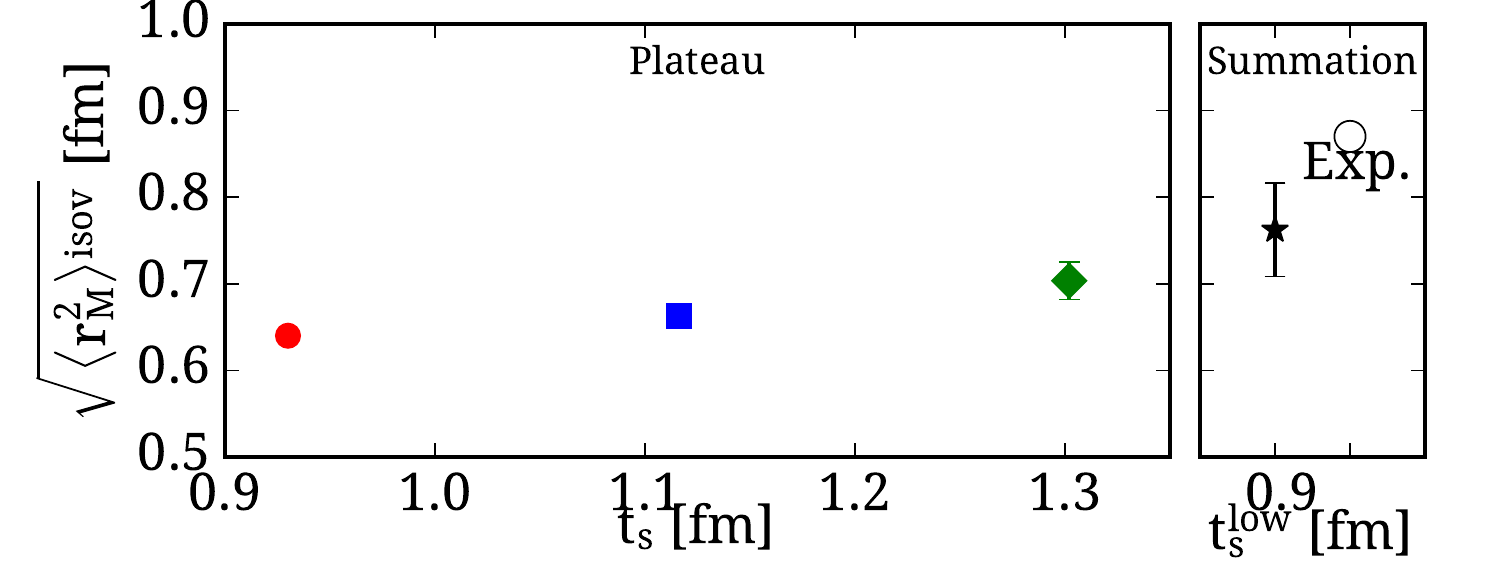}
  \end{minipage}
  \caption{Results for isovector $\langle r_E^2\rangle$ (left) and
    $\langle r_M^2\rangle$ (bottom right) and $G_M(0)$ (top right)
    from dipole fits. Fits to results using the plateau method are
    shown with the symbol notation of Fig.~\protect\ref{fig:em ffs}. For the
    summation method we fit all available $t_s$ to obtain the filled
    asterisk and starting from 1.1~fm using the open asterisk. The
    open circles are the experimental result from
    Ref.~\protect\cite{Mohr:2015ccw} while the open square from
    Ref.~\protect\cite{Pohl:2010zza}.}
  \label{fig:em fits}
\end{figure}

The electric and magnetic radii are related to the slope of the form
factors at $Q^2=0$, namely: $\langle
r^2_{i}\rangle=-\frac{6}{G_i(0)}\partial G_i(Q^2)/\partial Q^2$,
$i=E,M$. We fit all sink-source separations and the summation method
to a dipole form $G_i(Q^2) = G_i(0)/(1+\frac{Q^2}{M_i^2})^2$ with
$\langle r^2_{i}\rangle=\frac{12}{M^2_i}$. For $G_E$ we fix $G_E(0)=1$
while for $G_M$ we allow $G_M(0)$ to vary. The results are shown in
Fig.~\ref{fig:em fits} where for both electric and magnetic radii we
see an increasing trend towards the experimentally determined values
as $t_s$ increases, while $G_M(0)$ shows mild dependence on $t_s$.

Our results for the isovector electric and magnetic charge radii are
compared to those of other recent lattice calculations in
Fig.~\ref{fig:radii comparisons}. We see that for both radii lattice
results agree within errors and are within at most 2-$\sigma$ to the
experimental values. With further improvement on systematic
uncertainties and with increased statistics, contacting experiment is
within reach for these quantities.

\section{Summary and conclusions}
The isovector axial and electromagnetic form factors of the nucleon
have been calculated on a lattice with physical pion mass at multiple
sink-source separations up to $\sim$1.7~fm and for $\mathcal{O}(10^5)$
statistics. We find that excited states increase the axial mass and at
separations beyond 1.3~fm our result agrees with experimental
measurements.

\begin{figure}
  \begin{minipage}{0.3\linewidth}
    \includegraphics[width=\linewidth]{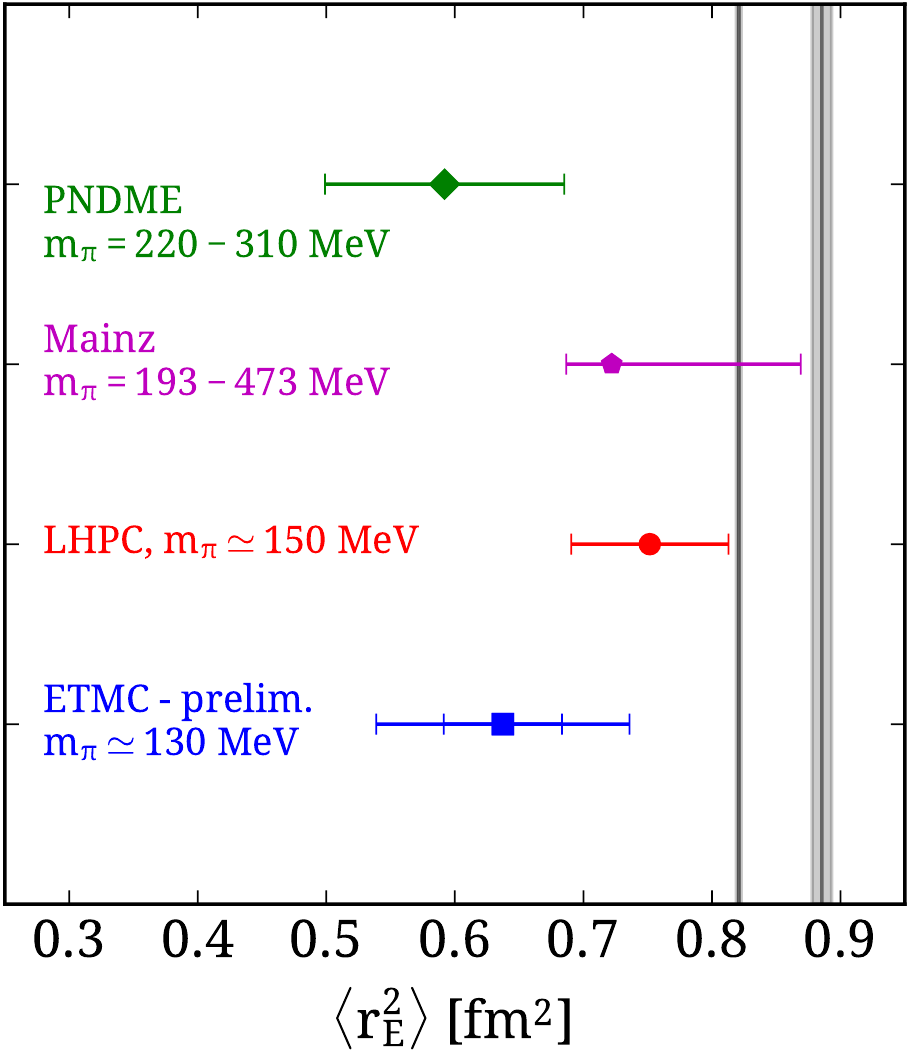}
  \end{minipage}
  \hfill
  \begin{minipage}{0.3\linewidth}
    \includegraphics[width=\linewidth]{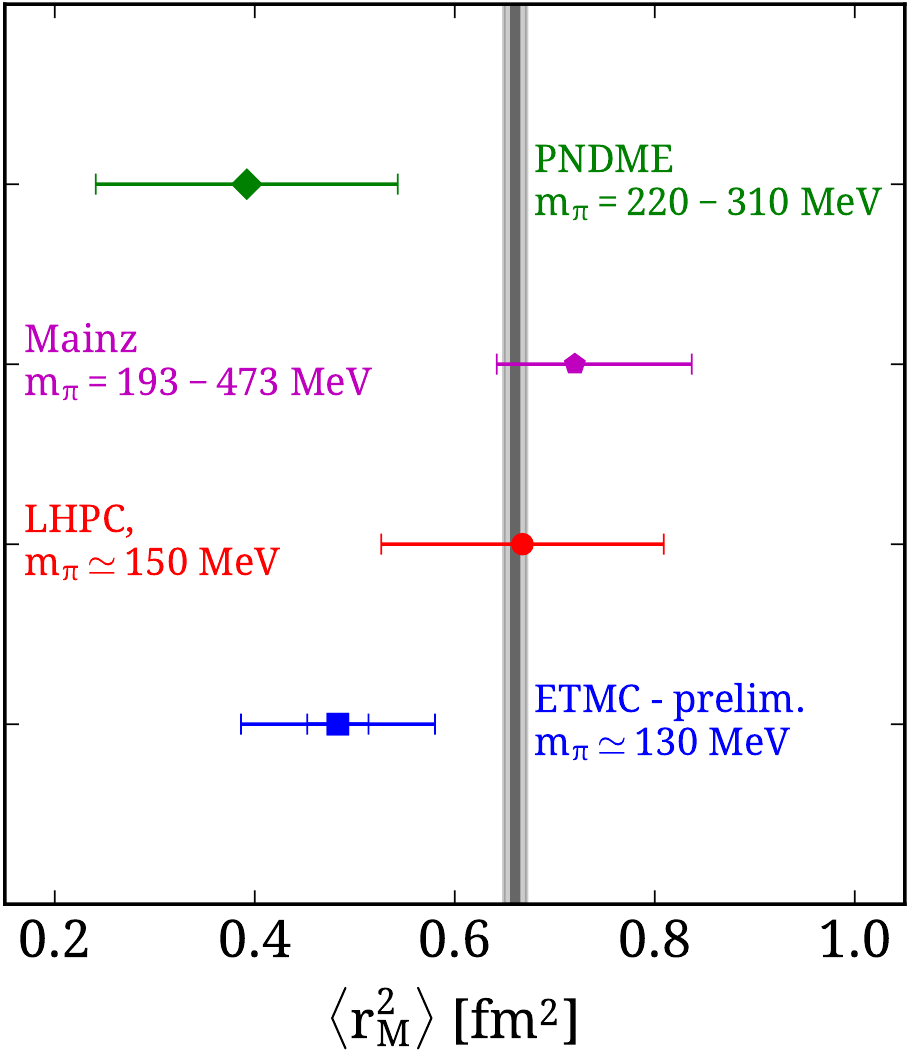}
  \end{minipage}
  \hfill
  \begin{minipage}{0.348\linewidth}
    \includegraphics[width=\linewidth]{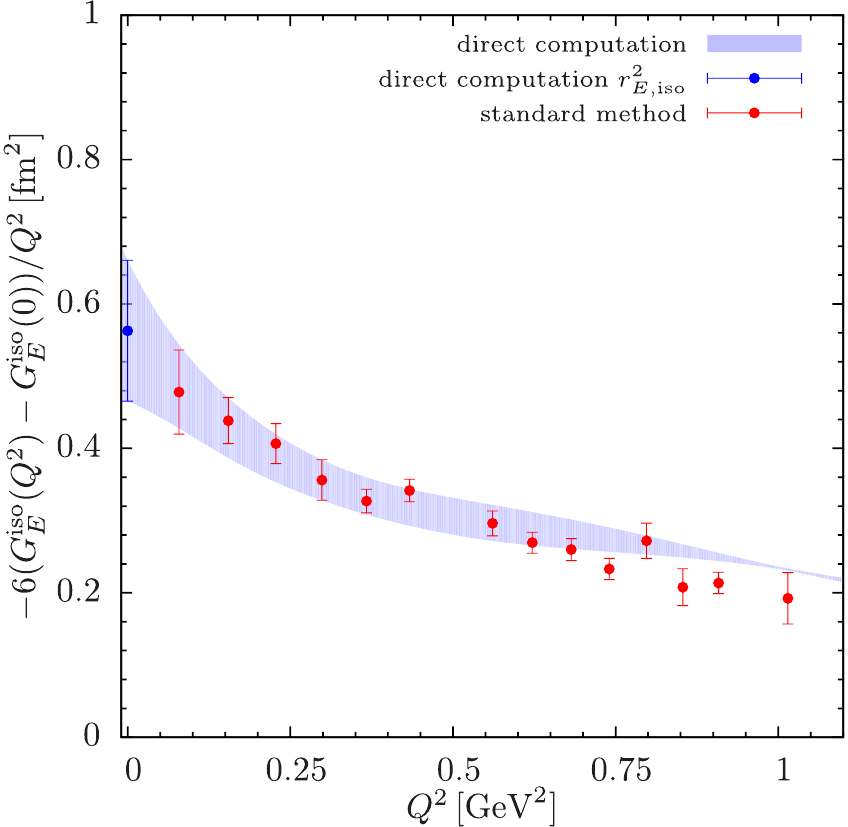}\\[-1.4ex]
  \end{minipage}
  \caption{Our results for the isovector $\langle r_E^2\rangle$ (left)
    and $\langle r_M^2\rangle$ (center), shown with blue squares using
    the plateau method for the largest $t_s$ in each case. The smaller
    error-bar indicates the statistical error while the larger error
    includes the systematic uncertainty when considering the summation
    method. We compare to recent lattice calculations:
    PNDME~\protect\cite{Bhattacharya:2013ehc} (green diamonds),
    Mainz~\protect\cite{Capitani:2015sba} (magenta pentagons) and
    LHPC~\protect\cite{Green:2014xba} (red circles). The vertical lines show
    the experimental values also shown in Fig.~\protect\ref{fig:em fits}. In
    the right panel we show a preliminary result of using the position
    space method of Ref.~\protect\cite{Alexandrou:2016rbj} for determining the
    slope of $G_E(Q^2)$ for the separation $t_s\simeq$1.7~fm.}
  \label{fig:radii comparisons}
\end{figure}

The electric and magnetic charge radii show similar behavior,
approaching the experimental values with increasing $t_s$. For
$G_M(0)$, the value at $Q^2=0$ is underestimated with mild excited
state dependence. Calculations at a larger volume, with access to
finer momenta, are being carried out to asses the effect on $G_M(0)$.

Recent methods for fitting form factors with no model assumption of
the their $Q^2$ dependence allow for further assessment of systematic
uncertainties. In the right panel of Fig.~\ref{fig:radii comparisons}
we show the result of applying the position space method of
Ref.~\cite{Alexandrou:2016rbj}, originally applied for $G_M(0)$, to
obtain $\langle r^2_E\rangle$ at $t_s=1.7$~fm. At all separations we
obtain results for $\langle r^2_E\rangle$ consistent with what is
obtained by the dipole fits shown in Fig.~\ref{fig:em fits}. Such
methods can benefit from finer momenta using larger lattice volumes,
as well as from reduced errors at larger $Q^2$ using appropriate
momentum-dependent smearing as in Ref.~\cite{Bali:2016lva}, both
avenues which are currently being explored.

\textit{Acknowledgments:} Results were obtained using Jureca, via NIC
allocation ECY00, HazelHen at HLRS and SuperMUC at LRZ via Gauss
allocations with ids 44066 and 10862 and Piz Daint at CSCS via
projects with ids s540 and s625. We thank the staff of these centers
for access to the computational resources and for their support.

\bibliographystyle{h-physrev}
\bibliography{refs.bib}
\end{document}